\documentclass[twocolumn,English,aps,pra,10pt,superscriptaddress,floatfix]{revtex4-2}
\usepackage{times}
\usepackage{graphicx}
\usepackage{amssymb}
\usepackage{bm}
\usepackage{amssymb,amsfonts,amsmath,amsbsy,bm,t1enc,latexsym}
\usepackage{float}
\usepackage[colorlinks=true,citecolor=blue,linkcolor=magenta]{hyperref}
\usepackage[markup=blue, authormarkupposition=left]{changes}
\usepackage{url}
\usepackage{xspace} 
\usepackage{siunitx}
\DeclareSIUnit \dbc {dBc}
\usepackage{soul}
\usepackage{changes}
\usepackage{cancel}
\usepackage{times}
\usepackage{soul}
\usepackage{amsmath}
\usepackage{hyperref}

\newcommand{\fref}[1]{Fig.~\ref{#1}}

\newcommand{\LN}[0]{$\mathrm{LiNbO}_3$} 
\newcommand{\LT}[0]{$\mathrm{LiTaO}_3$} 
 
\newcommand{\SiO}[0]{$\mathrm{SiO}_2$}

\begin{document}

\title{Ultrabroadband thin-film lithium tantalate modulator for high-speed communications}




\author{Chengli Wang}\thanks{These authors contributed equally.}
\affiliation{Institute of Physics, Swiss Federal Institute of Technology Lausanne (EPFL), CH-1015 Lausanne, Switzerland}
\affiliation{Center of Quantum Science and Engineering, EPFL, CH-1015 Lausanne, Switzerland}

\author{Dengyang Fang}\thanks{These authors contributed equally.}
\affiliation{Institute of Photonics and Quantum Electronics (IPQ), Karlsruhe Institute of Technology (KIT), 76131 Karlsruhe, Germany}

\author{Junyin Zhang}\thanks{These authors contributed equally.}
\affiliation{Institute of Physics, Swiss Federal Institute of Technology Lausanne (EPFL), CH-1015 Lausanne, Switzerland}
\affiliation{Center of Quantum Science and Engineering, EPFL, CH-1015 Lausanne, Switzerland}

\author{Alexander Kotz}
\affiliation{Institute of Photonics and Quantum Electronics (IPQ), Karlsruhe Institute of Technology (KIT), 76131 Karlsruhe, Germany}

\author{Grigory Lihachev}
\affiliation{Institute of Physics, Swiss Federal Institute of Technology Lausanne (EPFL), CH-1015 Lausanne, Switzerland}
\affiliation{Center of Quantum Science and Engineering, EPFL, CH-1015 Lausanne, Switzerland}

\author{Mikhail Churaev}
\affiliation{Institute of Physics, Swiss Federal Institute of Technology Lausanne (EPFL), CH-1015 Lausanne, Switzerland}
\affiliation{Center of Quantum Science and Engineering, EPFL, CH-1015 Lausanne, Switzerland}

\author{Zihan Li}
\affiliation{Institute of Physics, Swiss Federal Institute of Technology Lausanne (EPFL), CH-1015 Lausanne, Switzerland}
\affiliation{Center of Quantum Science and Engineering, EPFL, CH-1015 Lausanne, Switzerland}

\author{Adrian Schwarzenberger}
\affiliation{Institute of Photonics and Quantum Electronics (IPQ), Karlsruhe Institute of Technology (KIT), 76131 Karlsruhe, Germany}

\author{Xin Ou}
\email[]{ouxin@mail.sim.ac.cn}
\affiliation{National Key Laboratory of Materials for Integrated Circuits, Shanghai Institute of Microsystem and Information Technology, Chinese Academy of Sciences, Shanghai, China}

\author{Christian Koos}
\email[]{christian.koos@kit.edu}
\affiliation{Institute of Photonics and Quantum Electronics (IPQ), Karlsruhe Institute of Technology (KIT), 76131 Karlsruhe, Germany}

\author{Tobias Kippenberg}
\email[]{tobias.kippenberg@epfl.ch}
\affiliation{Institute of Physics, Swiss Federal Institute of Technology Lausanne (EPFL), CH-1015 Lausanne, Switzerland}
\affiliation{Center of Quantum Science and Engineering, EPFL, CH-1015 Lausanne, Switzerland}

\maketitle


\textbf{
The continuous growth of global data traffic over the past three decades, along with advances in disaggregated computing architectures, presents significant challenges for optical transceivers in communication networks and high-performance computing systems. Specifically, there is a growing need to significantly increase data rates while reducing energy consumption and cost. High-performance optical modulators based on materials such as InP, thin-film lithium niobate (\LN), or plasmonics have been developed, with \LN~excelling in high-speed and low-voltage modulation. Nonetheless, the widespread industrial adoption of thin film \LN~remains compounded by the rather high cost of the underlying 'on insulator' substrates -- in sharp contrast to silicon photonics, which can benefit from strong synergies with high-volume applications in conventional microelectronics. Here, we demonstrate an integrated 110 GHz modulator using thin-film lithium tantalate (\LT) --- a material platform that is already commercially used for millimeter-wave filters and that can hence build upon technological and economic synergies with existing high-volume applications to offer  scalable low-cost manufacturing.  We show that the \LT~ photonic integrated circuit based modulator can support 176 GBd PAM8 transmission at net data rates exceeding 400~$\mathrm{Gbit/s}$, while exhibiting a lower bias drift compared to \LN. Moreover, we show that using silver electrodes can reduce microwave losses compared to previously employed gold electrodes. Our demonstration positions \LT~modulator as a novel and highly promising integration platform for next-generation high-speed, energy-efficient, and cost-effective transceivers.}\\




The relentless increase in global data traffic, driven by the widespread use of novel technologies such as 5G and artificial intelligence (AI), has created significant challenges for transceivers at all levels of optical networks~\cite{tauber2023role, winzer2018fiber}. These challenges include increased transmission rates along with reduced energy consumption and costs. Over the previous years, silicon photonics has been widely deployed in the optical transceiver market, mainly driven by the cost-efficiency of the underlying silicon-on-insulator (SOI) substrates and the amenability to high-volume production of photonic integrated circuits (PIC) using technically mature CMOS processes \cite{xu2005micrometre, sun2015single, thomson2016roadmap}. However, on a technical level, silicon photonic electro-optic modulators have to rely on free-carrier dispersion to overcome the intrinsic lack of Pockels-type nonlinearities on bulk silicon. This approach is currently reaching its physical  limits~\cite{zhang2022ab, thomson2016roadmap} in terms of bandwidth, power consumption, and impairments such as free-carrier absorption and modulation nonlinearity, especially given the future demand for highly efficient transceivers that offer line rates of 1.6~Tbit/s or more~\cite{tauber2023role}.

Apart from silicon, substantial efforts have been made towards developing high-performance optical modulators across various material platforms, such as indium phosphide (InP)~\cite{ogiso201980}, thin-film lithium niobate~\cite{zhang2021integrated, he2019high, berikaa2023tfln}, plasmonic and silicon-organic hybrid (SOH)~\cite{kulmer2024single, kieninger_silicon-organic_2020,freude_high-performance_2024} and other platforms \cite{abel2019large, phare2015graphene, han2017efficient}. 
Ferroelectric thin-film lithium niobate platform offers low optical and microwave losses, high optical power handling, as well as high Pockels coefficient, while enabling linear high-speed modulation at low voltage levels without performance degradation over time~\cite{zhang2021integrated, berikaa2023tfln, wang2018integrated}. 
However, despite tremendous research progress in device design and demonstrations, it is still an open question whether thin-film \LN~modulators can achieve market penetration on the same scale as silicon photonics does today. One reason is the high cost of \LN-on-insulator (LNOI) wafers, which is a major obstacle towards adoption of the technology in cost-sensitive transceiver markets. Specifically, LNOI substrate technology cannot rely on any high-volume applications outside photonics --- unlike silicon photonics and the underlying silicon-on-insulator (SOI) substrates, which were driven by significant investments into CMOS technology over the previous three decades. 

\begin{figure*}[htbp]
	\centering
	\includegraphics[width=0.98\linewidth]{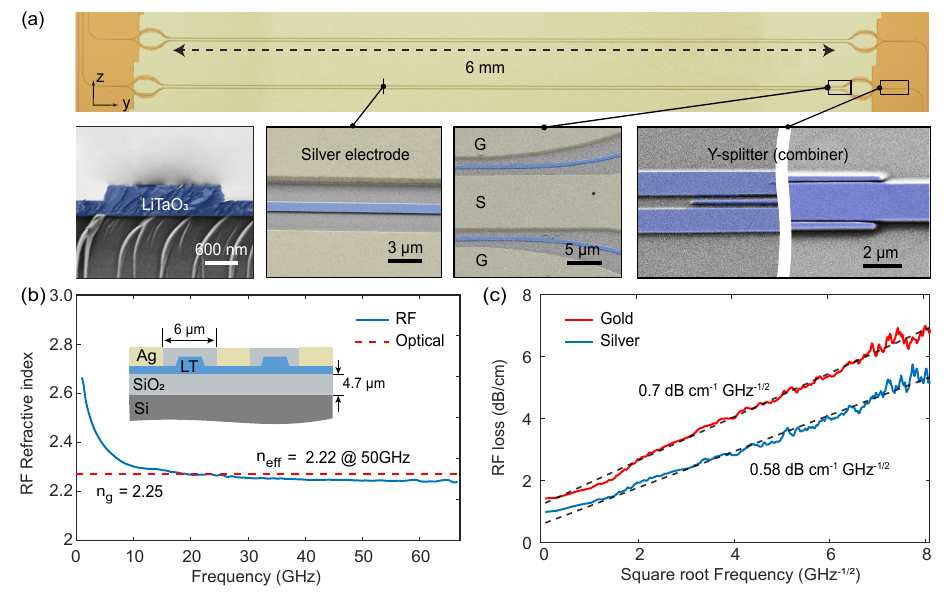}
	\caption{\textbf{Thin-film Lithium Tantalate electro-optic Mach-Zender modulator.} 
	(a) Microscope image of a fabricated \LT~modulator. Inset scanning electron microphotography (SEM)images show the key components of the \LT~modulator, including the cross-sectional \LT~waveguide (blue), the silver electrode (yellow), the ground-single-ground (GSG) configuration and the Y-spliter.  (b) Phase matching between the optical and microwave waves. The simulated group index $n_g$ = 2.25 of the optical \LT~waveguide is marked in a dotted line. The measured RF phase index is $n_{\mathrm{eff}}$ = 2.22 at 50 GHz. Inset: Cross-section of MZM covered with \SiO~cladding. Parameters are: electrode thickness: \SI{800}{\nano\meter}, cladding thickness: \SI{1.4}{\micro\meter}, \LT~thickness (half-etched): \SI{600}{\nano\meter}, BOX layer: \SI{4.7}{\micro\meter}, electrode gap: \SI{6}{\micro\meter}, waveguide width: \SI{1.2}{\micro\meter}, signal electrode width: \SI{19}{\micro\meter}. (c) Measured radiofrequency (RF) losses on a square-root frequency axis.
	}
	\label{fig1}
\end{figure*}

In contrast to that, another ferroelectric material, lithium tantalate, has achieved mass production due to its wide application in RF filters for 5G \cite{ballandras2019new}. This already existing substantial fabrication volume has allowed to mature the technology and to address cost challenges when adopting \LT-on-insulator (LTOI) as a platform for optical modulators. Recently, the first low-loss \LT~PICs have been demonstrated using diamond-like carbon (DLC) as a hardmask~\cite{wang2024lithium}, and equivalent or even superior performance of \LT~compared to \LN~has been demonstrated. Specifically, the optical birefringence \LT~\((\Delta n = n_{e} - n_{o} = 0.004\)) is $\times$17 lower compared to \LN~\((\Delta n = -0.07\)), which facilitates the design and development of compact devices with tight bends. 
Moreover, \LT~has a significantly higher optical damage threshold~\cite{yan2020high} and weaker photorefractive effect~\cite{wang2024lithium}, which helps mitigate DC bias drift problems that are commonly observed in \LN~modulators~\cite{holzgrafe2024relaxation, zhang2021integrated, yu2024tunable}. Given these advantages, \LT~is expected to match or even surpass \LN~in performance for various photonic devices --- at reduced cost. However, to date, the core component of optical communications, the ultra-high speed electro-optic modulator, has not yet been demonstrated in LTOI.

Here, we report on a high-speed \LT~Mach-Zehnder modulator (MZM) with a measured electro-optic 3 dB bandwidth of approximately 110 GHz and a half-wave voltage-length product of 2.8 V·cm. We achieve these performance parameters by engineering the microwave and photonic circuits and by applying high-conductivity electrodes to  simultaneously achieve low microwave losses and group-velocity-matching. We further conducted data communication experiments using the LTOI modulator and demonstrated PAM8 transmission at a symbol rate of 176 GBd, achieving a single-carrier net data rate of more than 400 Gbit/s with a bit-error ratio (BER) below the threshold for soft-decision forward-error correction (SD-FEC) with 25\% coding overhead. Additionally, we show that the LTOI device exhibits a lower bias drift compared to an \LN~integrated modulator.

\begin{figure*}[!htbp]
	\centering
	\includegraphics[width=0.95\linewidth]{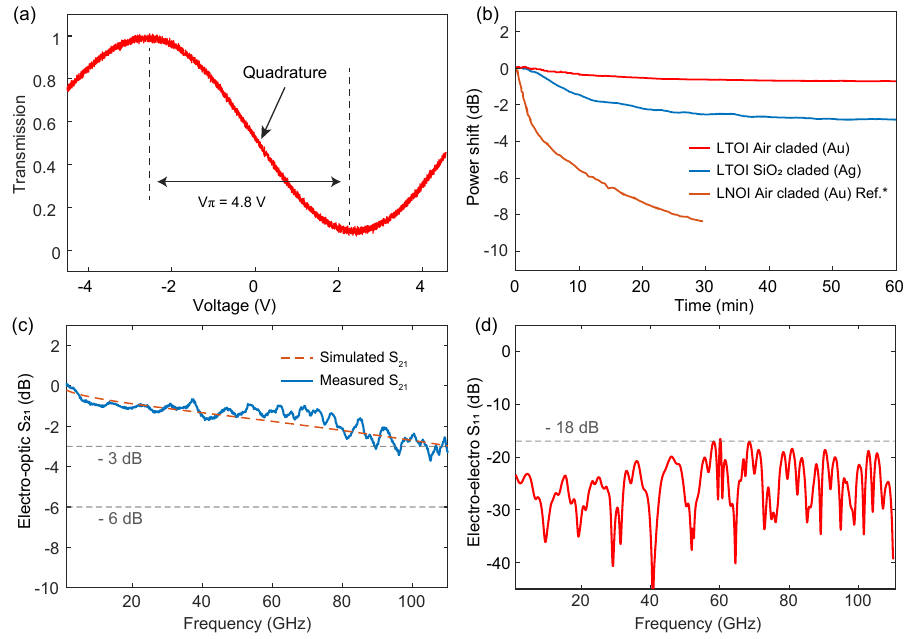}
	\caption{\textbf{ Electro-optic performance of the \LT~Mach-Zender modulator.} 
     (a) Measured DC and extinction for a 6 mm long modulator. (b) Dynamics of relative output intensity when bias set to quadrature at quadrature as a function of the operating time in the \LT~MZM and comparison with thin-film \LN~modulators. The bias drift data of \LN~in yellow curve  is from Ref \cite{xu2020high}.(c) EO bandwidth ($\mathrm{S}_{21}$) and (d) electrical reflection ($\mathrm{S}_{11}$) of the \LT~modulator, showing a high 3-dB bandwidth at around 110 GHz. The simulated EO response is calculated from the electro-electro measurement. 
	}
	\label{fig2}
\end{figure*}

\fref{fig1}(a) shows a fabricated \LT~MZM composed of two 50:50 adiabatic Y-splitters at either end and a push-pull optical waveguide phase shifter pair with a length of \SI{6}{\milli\meter}. In our work we use an x-cut single crystalline \LT~thin film, which was fabricated from a optical-grade bulk \LT~wafer by using ion-cutting and wafer bonding methods. 
The fabricated LTOI wafer stack consists of a 600 nm thin-film \LT, a \SI{4.7}{\micro\meter} thick thermal silicon dioxide, and a \SI{525}{\micro\meter} thick high-resistivity silicon carrier wafer.
The device fabrication process chosen in this work was based on die level processing and involved three main steps: 1) electron beam patterning; 2) ion-beam dry-etching and KOH wet-etching; 3) lift-off of the coplanar waveguide (CPW) electrodes. 
The performance of a high-speed traveling-wave electro-optic modulators based on ferroelectric thin films depends on the impedance of the RF waveguide, the loss of the RF signal, and the matching of the propagation speeds of the optical and RF signals. The relevant theory is well-developed in the integrated modulator field \cite{alferness1984velocity, wang2018integrated, zhang2021integrated, zhu2021integrated}. To make a trade-off between decreasing the RF loss and the modulation efficiency, the \LT~waveguide width is chosen to \SI{1.2} {\micro\meter} and the gap between the waveguide sidewalls and the electrode is \SI{2.4} {\micro\meter} on each side. By designing the thickness of each layers of the LTOI wafer, we can achieve good velocity matching between the RF and optical waves.
 \fref{fig1}(b) presents the measured RF refractive index of a \SiO~cladded 600 mm long MZM devices, showing well-matched velocity with $n_{\mathrm{eff}}$ = 2.22 at 50 GHz and $n_{g}$ = 2.25. The detailed cross-section structure and the layer thicknesses are shown in the inset of \fref{fig1}(b).

For the RF attenuation, instead of the commonly used gold CPW in \LN~modulator \cite{he2019high, wang2018integrated, xu2020high}, silver is deposited with an 800 nm thickness as the CPW electrodes in this work. Although gold is widely employed due to its low resistivity ($\rho_\mathrm{Au} = 2.2\times10^{-6}~\Omega~cm$) and excellent process stability, silver is showing an even lower resistivity ($\rho_\mathrm{Ag} = 1.55\times10^{-6}~\Omega~cm$) \cite{matula1979electrical}. To evaluate the electrical losses of gold and silver electrodes, we fabricated CPWs using the same device parameters with both silver and gold. \fref{fig1}(c) shows the microwave loss measurement result of the silver and gold CPWs using a 67 GHz vector network analyzer (VNA). The results indicate that silver CPWs exhibit better linear loss due to their lower resistivity, with $\mathrm{\alpha_{RF, Ag}} = 0.58~\mathrm{dB~cm}^{-1}\mathrm{GHz}^{-1/2}$ compared to gold, which has $\mathrm{\alpha_{RF, Au}} = 0.77~\mathrm{dB~cm}^{-1}\mathrm{GHz}^{-1/2}$ (Figure 1(c)). The square root dependence of the loss in Figure 2(b) also confirms the assumption that ohmic loss is the dominant source of RF attenuation. Additionally, silver demonstrated good stability over a one-month exposure test, showing no performance degradation. This indicates that silver is a promising CPW electrode material with low ohmic loss, good stability, and greater economic efficiency.

The static $V_\pi$ characterization is performed first. 
Light at 1550 nm is coupled in and out using two lensed fibers with a coupling loss of about 7 dB per facet. The majority of the coupling loss results the large mismatch between the mode sizes and mode indices of the optical fiber and the partially-etched \LT~rib waveguide. This occurs because tapering the rib waveguide pushes the optical mode into the slab region, which can be optimized by using a mode size converter \cite{ying2021low}.
The low propagation loss performance of \LT~has been demonstrated in our previous work \cite{wang2024lithium}. We fabricated additional optical \LT~waveguides without CPW to investigate the metal absorption loss in the current devices. The measured results showed no measurable optical loss difference between the reference waveguides and the real device waveguides, indicating that the absorption loss of the electrodes in our current design length is negligible, consistent with the simulation, which showed an electrode absorption loss of 0.03 dB/cm. For a 6 mm MZM with 100 Hz triangular voltage sweeps, the measured $V_\pi$ is \SI{4.8}{\volt} at 1550 nm, corresponding to $V_\pi L$ of \SI{2.8}{\volt\cdot\centi\meter}. The $V_\pi L$ increase relative to the previous result \cite{wang2024lithium}, can be attributed to the enlarged gap between the ground and signal electrode. By refining the design parameters, specifically through extending the length of the modulator and narrowing the electrode gap, or by using a substrate with lower RF absorption losses like quartz \cite{xu2022dual}, it is feasible to achieve high-bandwidth and low-voltage operations on \LT. 

Next, we study the bias drift of the modulator. Although thin film \LN~modulator has been well developed, the issue of EO relaxation in \LN~poses a considerable challenge for long-timescale applications. The causes of EO relaxation in \LN~are investigated to be related to factors such as interface defects and the photorefractive effect \cite{xu2021mitigating, holzgrafe2024relaxation}.
\LT~has been demonstrated a weaker photorefractive (PR) effect and lower drift phenomenon compared to \LN~in optical microresonators \cite{wang2024lithium, yu2024tunable}.
In the fabricated \LT~MZM, we set a quadrature bias and measured the DC drift over 60 minutes.
We use two types of \LT~MZMs devices: one with air cladding and gold as the CPW material, and another with \SiO~cladded and silver as the CPW material. We observed that the DC drift behavior differed between these two devices. The MZM with \SiO-cladding and silver as electrode a drift of about 3 dB, while the device with air cladding and gold as the CPW material had a drift of only 1 dB. The difference in drift between these two devices might be due to the presence of more traps at the \SiO~cladding interface, enhancing the PR effect \cite{xu2021mitigating}. The different CPW materials might also contribute to the variation in DC drift behavior, which requires further investigation.
However, in both cases, the DC drift in the LTOI is much smaller than that observed 8 dB drift in \LN~\cite{xu2020high}, indicating better DC stability in \LT~MZM.

We then characterized the small-signal electro-optic (EO) bandwidth of the fabricated devices. The measured EO response of the \LT~MZM in \fref{fig2}(c) shows a flat response over the frequency range with a roll-off around 3 dB from 10 MHz to 110 GHz. \fref{fig2}(d) shows the measured electrical reflection of the silver CPW below -18 dB at up to 110 GHz revealing that the good impedance matching was achieved and the reflection levels are sufficiently low for practical applications. 

\begin{figure*}[htbp]
	\centering
	\includegraphics[width=0.9\linewidth]{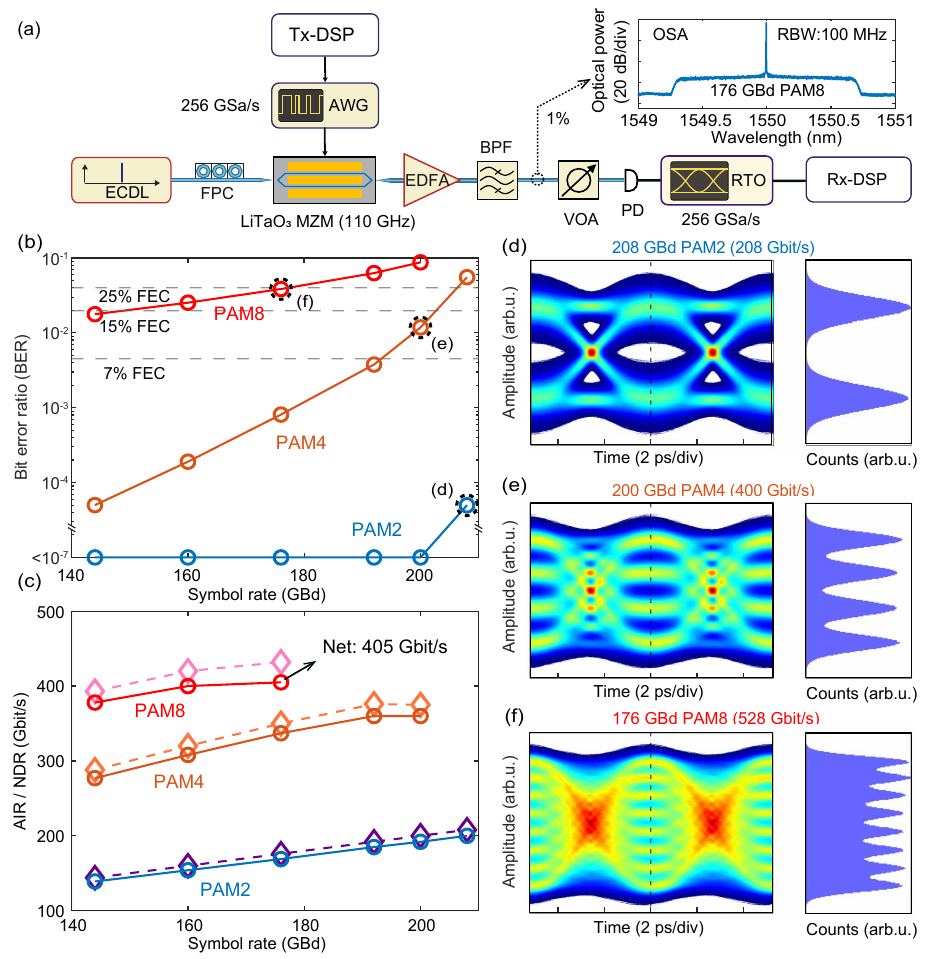}
	\caption{\textbf{Schematic and results of the intensity-modulation direct detection (IMDD) experiment employing the thin-film \LT~Mach-Zender modulator.} 
(a) Experimental setup: an external cavity laser (ECL) serves as a light source, and a polarization controller (PC) is used to tune the polarization of the light. The optical input/output coupling to the 110 GHz \LT~MZM relies on a pair of lensed fibers. The drive signals are synthesized by transmitter digital signal processing (Tx-DSP) and generated by an arbitrary waveform generator (AWG). The modulated light is amplified by an erbium-doped fiber amplifier (EDFA), and the out-of-band amplified spontaneous emission (ASE) noise is suppressed by a tunable bandpass filter (BPF). 99\% of the amplified light enters a variable optical attenuator (VOA) before being detected by a photodiode (PD). A high-speed real-time oscilloscope (RTO) samples the resulting signal, which is processed offline by receiver DSP (Rx-DSP). 1\% of the amplified light after the EDFA is sent to an optical spectrum analyzer (OSA) for monitoring. The inset 1 shows an exemplary optical spectrum for a 176 GBd PAM8 signal. 
(b) Measured bit error ratios (BER) as a function of symbol rates for PAM8 (red), PAM4 (green), and PAM2 (blue) signals. Horizontal black dashed lines indicate the thresholds for 25\%, 15\% soft-decision and for 7\% hard-decision forward error correction (FEC). 
(c) Extracted available information rates (AIR, dashed lines) and associated net data rates (NDR, solid lines) for measurements with BER values below 25\% SD-FEC limit. The yellow star marks the highest NDR of 405 Gbit/s achieved by using a PAM8 signal at a symbol rate of 176 GBd.
(d)-(f) Eye diagrams and associated histograms taken in the center of the symbol slot (vertical dashed line) for the circled data points marked in Subfigure (b). 
	}
	\label{fig3}
\end{figure*}

To demonstrate the outstanding performance of the designed and fabricated thin-film \LT~MZM, a high-speed data communication experiment was conducted. The underlying setup is depicted in \fref{fig3}(a). The optical carrier is provided by an external-cavity laser (ECL), which, after a fiber polarization controller (PC), is coupled into the MZM through a lensed fiber. The electrical drive signal is generated by a high-speed arbitrary waveform generator (AWG, M8199B, Keysight Technologies Inc., CA, USA) and fed to the MZM via a 20 cm-long RF cable and a 110 GHz RF probe. At the transmitter, we use digital signal processing (Tx-DSP) to generate various pulse-amplitude modulation (PAM) signals based on pseudo-random bit sequences (PRBS) and root-raised cosine (RRC) pulse-shaping filters with a roll-off of $\rho=0.05$. We account for the frequency-dependent RF loss in the cable by applying a linear minimum-mean-square-error (MMSE) predistortion. A second 110 GHz RF probe with an attached bias-T followed by a 50 Ohm resistor is used to terminate the device and to set the MZM bias point to quadrature point for intensity modulation. The intensity-modulated optical signal generated by the MZM is coupled out through a second lensed fiber and amplified by an erbium-doped fiber amplifier (EDFA). Unwanted out-of-band amplified spontaneous-emission (ASE) noise of the EDFA is suppressed by a tunable bandpass filter (BPF, Koshin Kogaku Co. Ltd., Kanagawa, Japan). A tap separates 1\% of the light after the BPF to monitor the optical spectrum of the signal by an optical spectrum analyzer (OSA, Apex Technologies, FR). The remaining 99\% of the light is passed through a variable optical attenuator (VOA, LTB-1, EXFO Inc., CA) to adjust for an optical output power of 10 dBm, and the signal is then sent to the subsequent high-speed 90 GHz photodiode (PD, Finisar Corp., CA, USA) for direct detection. The received electrical signal is digitized by a real-time oscilloscope (RTO, UXR 1004A, Keysight Technologies Inc., CA, USA) with an analogue bandwidth of 100 GHz and a sampling rate of 256 GSa/s. The data is finally extracted by offline receiver DSP (Rx-DSP), which contains resampling to 2 Sa/sym, timing recovery, linear Sato equalization, and an additional decision-directed least-mean-square (DD-LMS) equalizer. 

We use the setup shown in \fref{fig3} to generate and receive PAM2, PAM4 and PAM8 data signals with symbol rates ranging from 144 GBd to 208 GBd in an optical back-to-back configuration. Inset in \fref{fig3} (a) shows an exemplary optical spectrum of 176 GBd PAM8 measured by the OSA. After the Rx-DSP, the bit error ratios (BER) of the various PAM signals are calculated and plotted as a function of symbol rate, see \fref{fig3} (b).
The limits for soft-decision forward-error correction (SD-FEC) with 25\% and 15\% overhead and for hard-decision forward error correction (HD-FEC) with 7\% overhead are indicated by horizontal black dashed lines. The results indicate that we can transmit PAM8 signal with a symbol rate of 176 GBd, while the measured BER of $3.8 \times 10^{-2}$ is still below the threshold of 25\% SD-FEC limit. For 200 GBd PAM4 and 208 GBd PAM2, the achieved BERs are well below the threshold of 15\% SD-FEC and 7\% HD-FEC limits, respectively. The eye diagrams and histograms of the circled data points are depicted in \fref{fig3} (d). The histograms show the display of the signal amplitudes sampled at the center of the eye diagrams indicated by vertical dashed lines. We then calculate achievable information rates (AIR) by multiplying the symbol rates with the normalized mutual information (NGMI), where the NGMI values are calculated for each measurement based on log-likelihood ratios (LLR) by using an additive white Gaussian noise (AWGN) channel model. The corresponding results are displayed by curves in dashed lines in \fref{fig3} (b). In the same graph, we further plot the associated net data rates (NDR) for measurements with BER values below the 25\% SD-FEC limit in solid lines. The NDRs are obtained by using the respective code rate associated with the NGMI threshold, as measured in the publication~\cite{hu2022ultrahigh}, multiplied with the symbol rate. As a result, the highest AIR of 432 Gbit/s is achieved by using PAM8 signal at a symbol rate of 176 GBd. Our first proof-of-concept LTOI MZM can hence achieve a single-carrier net data rate of 405 Gbit/s, which is among the highest values so far achieved for MZM~\cite{berikaa2023tfln, kulmer2024single}, and which underlines the outstanding potential of the technology. Note that in this IMDD demonstration, only the most computing-efficient linear equalizations are employed, which leaves great room for further improvement in terms of net data rate by using more advanced probabilistic constellation shaping (PCS)~\cite{yamazaki2019net} or fast-than-Nyquist (FTN) coding~\cite{che2022higher} and by employing non-linear equalization techniques~\cite{liu2011intrachannel}. 


In conclusion, we have demonstrated the first high-speed LTOI based modulator, offering an electro-optic 3 dB bandwidth of 110 GHz. We prove the viability of the device by using it in IMDD transmission experiment, reaching a single-carrier net data rate of 405 Gbit/s, which is already on par with best-in-class LNOI and plasmonic platform devices~\cite{berikaa2023tfln, kulmer2024single}. 
We further show that using silver as an electrode material allows to reduce the microwave losses, and we find that LTOI devices offer a series of technical advantages with respect to their LNOI counterparts, such as increased DC bias stability. 
Using longer devices can further reduce the half-wave voltage to well below 1 V.
By combining the excellent signal fidelity with advanced in-phase/quadrature (I/Q) modulator design or more advanced signaling techniques and by incorporating polarization-division multiplexing, \LT~devices could offer line rates of 2 Tbit/s or more. 
With the mass production and availability of LTOI wafers, our results position LTOI as a highly promising integration platform for future electro-optic modulators that might outperform LNOI devices and that are key to future high-speed optical communication networks with increased throughput and efficiency. 



\section*{Acknowledgments}
The samples were fabricated in the EPFL Center of MicroNanoTechnology (CMi) and the Institute of Physics (IPHYS) cleanroom. T.J.K. acknowledges support from the Swiss National Science Foundation under grant agreement No. 216493 (HEROIC). This work was further supported by the BMBF project Open6GHub (no. 16KISK010), by the ERC Consolidator Grant TeraSHAPE (773248), by the DFG projects PACE (403188360) and GOSPEL (403187440) within the Priority Programme Electronic-Photonic Integrated Systems for Ultrafast Signal Processing (SPP 2111), and by the DFG Collaborative Research Centers (CRC) HyPERION (SFB 1527).

\section*{Author contributions} 
C.W. and X.O. fabricated the wafers.
C.W. and J.Z. designed the devices.
C.W. and J.Z. fabricated the devices.
D.F. and C.W. carried out the measurements.
D.F., C.W., and J.Z. analyzed the data.
C.W. and D.F. prepared the figures and wrote the manuscripts with contributions from all authors.
X.O.,C.K. and T.J.K. supervised the project.

\section*{Competing interests}
The authors declare no competing financial interests.

\section*{Data Availability Statement} The code and data used to produce the plots within this work will be released on the repository \textit{Zenodo} upon publication of this preprint.

\bibliography{refs}	
\bibliographystyle{naturemag}
\end{document}